\documentclass[prl, twocolumn, floatfix,showpacs,amsmath,amssymb]{revtex4}
\usepackage{bbm}
\usepackage{amsfonts}
\usepackage{graphicx}
\usepackage{color}
\usepackage{amsmath}
\usepackage{amssymb}
\usepackage{latexsym}
\usepackage{psfrag}

\begin{document}

\title{Interdependence of electroforming and hydrogen incorporation in nanoporous titanium dioxide}
\author{M. Strungaru\footnote{Present address: Alexandru Ioan Cuza University, CARPATH, Faculty of Physics, Blvd. Carol I, 11, 700506, Iasi, Romania}}
\author{M. Cerchez}
\author{S. Herbertz}
\author{T. Heinzel}\email{thomas.heinzel@hhu.de}
\affiliation{Solid State Physics Laboratory, Heinrich-Heine-Universit\"at D\"usseldorf, 40204 D\"usseldorf, Germany}
\author{M. El Achhab}
\author{K. Schierbaum}
\affiliation{Materials Science Laboratory, Heinrich-Heine-Universit\"at D\"usseldorf, 40204 D\"usseldorf, Germany}

\date{\today }

\begin{abstract}

It is shown that in nanoporous titanium dioxide films, sensitivity to atmospheric hydrogen exposure and electroforming can coexist and are interdependent. The devices work as conventional hydrogen sensors below a threshold electric field while above it, the well-known electroforming is observed. Offering hydrogen in this regime accelerates the electroforming process, and in addition to the usual reversible increase of the conductance in response to the hydrogen gas, an irreversible conductance decrease is superimposed. The behavior is interpreted in terms of a phenomenological model where current carrying, oxygen-deficient filaments with hydrogen-dependent conductivities form inside the $\mathrm{TiO_2}$ matrix.
\end{abstract}

\maketitle

Titanium dioxide is one of the most intensely studied metal oxides in present research.\cite{Diebold2003,Chen2007} Two important properties of $\mathrm{TiO_2}$ are memristive behavior, also known as electroforming, \cite{Strukov2008,Yang2008,Kwon2010,Szot2011} and strong response to exposure of various atmospheric gases.\cite{Yamamoto1980,Akbar1997,Chen2007,KimD2013} These effects have so far been studied only independently of each other. An interdependence would not only be of fundamental interest, but may also become relevant for applications. For example, electroforming may be one reason for drift effects in gas sensors, or the coupling strength of an artificial synapse \cite{Pickett2013} may be controllable by atmospheric additives. Here, we show that nanoporous titanium dioxide films \cite{Varghese2003,Paulose2006,Elachhab2014} show such an interdependence of hydrogen incorporation and memristive effects.

\begin{figure}[ht!]
\includegraphics[scale=1.0]{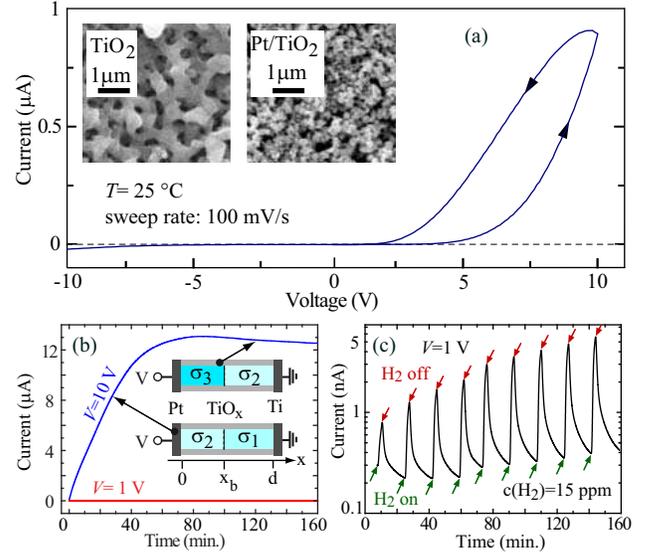}
\caption{(Color online). (a) $IV$ - measurement for a voltage loop in $\mathrm{N_2}$ atmosphere. The sweep direction is indicated by the arrows. Inset: scanning electron microscope images of the sample surface before (left) and after (right) coating with Pt. (b) Time dependence of $I$ at different voltages. Inset: sample geometry within the single filament model. Phases 1 to 3 are sketched before and after $I_{max}$ has been reached. (c) Current transients in response to $\mathrm{H_2}$ - pulses in $\mathrm{N_2}$ atmosphere for $V=1\,\mathrm{V}$. }
\label{TiOx_Fig1}
\end{figure}

\begin{figure}[ht!]
 \includegraphics[scale=1.0]{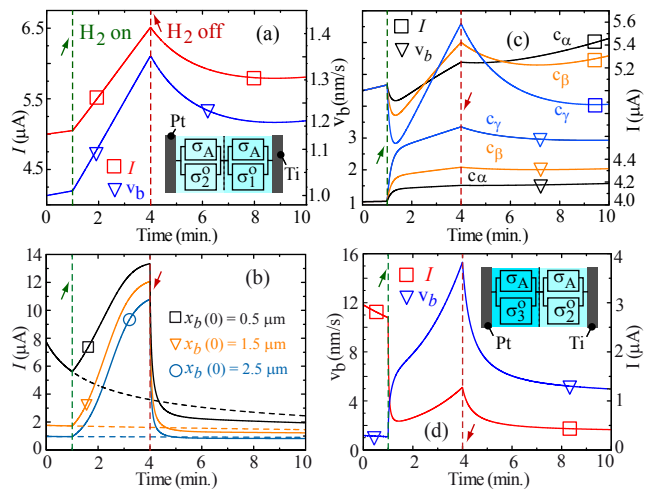}
 \caption{(Color online). (a) Simulation of $I(t)$ and $v_b(t)$ in a filament before $I(t)$ has reached $I_{max}$ (see the inset for the equivalent circuit), in response to $\mathrm{H_2}$ exposure of volume fraction $c_{\gamma}$. (b) Simulation of $I(t)$ and $x_b(t)$ as in (a) after $I_{max}$ has been reached, shown with (full lines) and without (dashed lines) hydrogen exposure. (c) Simulation as in (a) for hydrogen concentrations $c_{\alpha}<c_{\beta}<c_{\gamma}$, with mechanism B included. (d) Simulation of $I(t)$ and $v_b(t)$ as in (b) with mechanism B included. The inset shows the equivalent circuit used in ((b) and (d). The remaining simulation parameters are given in \onlinecite{Strungaru2015}. }
 \label{TiOx_Fig3}
 \end{figure}

A commercially available Ti foil (purity 99.6$\%$) is anodically oxidized at room temperature in a solution of $75\%$ $\mathrm{H_2SO_4}$ in DI water kept below $30\,^{\circ}\mathrm{C}$. The oxidation voltage was increased in steps up to $150\,\mathrm{V}$.\cite{Supplement} This produces a nanoporous oxide layer of $5\,\mathrm{\mu m}$ thickness, composed of a mixture of mostly rutile with a small admixture of anatase.\cite{Elachhab2014} The oxide surface is coated with Pt paste yielding a contact area of $12.5\,\mathrm{mm}^2$ area. The surface morphology before and after Pt coating is shown in the inset of Fig. \ref{TiOx_Fig1} (a). The sample is inserted in a measurement chamber with controlled atmosphere and initialized before each measurement sequence by exposing it to dry air for one hour at $T=180\,^{\circ}\mathrm{C}$.\cite{Supplement} Microscopic studies on single crystalline $\mathrm{TiO_2}(110)$ surfaces indicate a complicated interplay between surface defects and adsorbed $\mathrm{O_2}$. \cite{Wendt2005} In our system, exposure to $\mathrm{O_2}$ heals the point defects partially and leaves the sample in a weakly reduced state, $\mathrm{TiO_{2-\delta}}$. Afterwards, a voltage of $V=-30\,\mathrm{V}$ is applied to the Pt electrode with respect to the grounded Ti substrate for 12 hours, again in dry air, which causes the drift of oxygen ions ($\mathrm{O}^{2-}$) towards the Ti substrate,\cite{Yang2009} and of the donor-type oxygen vacancies \cite{Knauth1999} towards the Pt electrode. All subsequent experiments are carried out at room temperature, in the dark and in a nitrogen atmosphere (purity 99.999\%), to which molecular hydrogen gas ($\mathrm{H_2}$) is added via a gas flow controller after purging the lines to exclude contamination effects.\\
The sample initialized this way shows the typical response to atmospheric $\mathrm{H_2}$ \cite{Yamamoto1980,Harris1980,Varghese2003,Paulose2006} as well as electroforming, as illustrated in Fig. \ref{TiOx_Fig1}. Current-voltage ($IV$) - loops from $V=-10\,\mathrm{V}$ to $10\,\mathrm{V}$ and back are hysteretic, see Fig. \ref{TiOx_Fig1} (a), indicating electroforming. \cite{LeeS2014,Muenstermann2010} The absence of a hysteresis for negative voltages is due to the $\mathrm{Pt/TiO_{2-\delta}}$ Schottky barrier under reverse bias, which decreases the electric field in the bulk. According to the widely used model for electroforming in comparable systems,\cite{Kwon2010,Szot2011} filaments which predominantly carry the current form inside a $\mathrm{TiO_2}$ matrix. They are composed of a phase 1 of low conductivity $\sigma_1$ (smaller $\delta$), in series with a phase 2 with $\sigma_2 > \sigma_1$ (larger $\delta$), \cite{Strukov2008}, see the inset in Fig. \ref{TiOx_Fig1} (b). After the $IV$ loop, the phase boundary, located at $x_b$, thus resides close to the Pt electrode, and the sample has a low conductance.
\begin{figure}[ht!]
\includegraphics[scale=1.0]{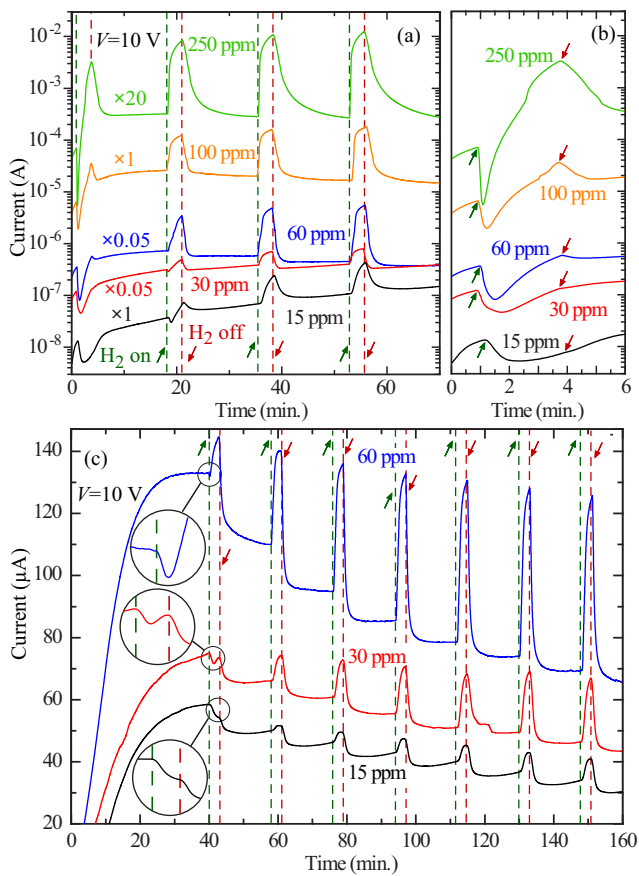}
\caption{(Color online). (a) $I(t)$ for $\mathrm{H_2}$ pulses of different volume fractions. Some traces have been scaled by factors as indicated to avoid overlaps. The times where hydrogen was turned on (off) is marked by the dashed lines and the upwards (downwards) pointing arrows. (b) Zoom-in of $I(t)$  shown in (a) for the first six minutes. (c) Response to $\mathrm{H_2}$ pulses close to and after current saturation. The traces for 15 ppm and 30 ppm have been offset by $-20\,\mathrm{\mu A}$ and $-10\,\mathrm{\mu A}$, respectively.}
\label{TiOx_Fig2}
\end{figure}
Afterwards, we apply positive voltages, such that the Schottky barrier is open and electroforming causes a current $I$ that increases with time $t$, see Fig. \ref{TiOx_Fig1} (b). $I$ is approximately constant for $V<2\,\mathrm{V}$, corresponding to electric fields $\varepsilon < 4\times 10^5\,\mathrm{V/m}$, for which electroforming is negligible. For $V=10\,\mathrm{V}$, $I(t)$ increases by two orders of magnitude as time proceeds. A maximum ($I_{max}$) is reached after 20 to 90 minutes, depending on the sweep, followed by a slight decrease. Pulsed measurements have been used to exclude Joule heating as a possible origin.\cite{Supplement} While the shape of $I(t)$ is reproducible, its amplitude varies up to a factor of 100. This indicates that a few filaments with distributed properties like their cross sectional area or $x_b$ are formed under identical external conditions. Within our model system composed of just one filament, the increase of $I(t)$ is caused by the expansion of phase 2 at the expense of phase 1. Saturation is reached for $x_b=d$. As the $\mathrm{O}^{2-}$ ions keep drifting towards the Pt electrode, phase 3 (small $\delta$) with a low conductivity $\sigma_3<\sigma_2$ is formed close to $x=0$. This phase grows in $x$-direction with time and causes the current to decrease. \cite{Supplement}\\
Below the threshold voltage, the current also responds to $\mathrm{H_2}$ as reported previously for similar structures,\cite{Varghese2003,Paulose2006,Cerchez2013} with a detection threshold of hydrogen volume fractions $c(\mathrm{H_2})<15\mathrm{ppm}$, see Fig. \ref{TiOx_Fig1} (c). Regarding the sensing mechanism, it has been established that $\mathrm{H_2}$ splits catalytically at the Pt film, the H atoms diffuse into the $\mathrm{TiO_{2-\delta}}$ \cite{Roland1997,Paulose2006,Aschauer2012}, form O-H groups with the oxygen of the host crystal \cite{Filippone2009,Bjorheim2010}  (interstitial hydrogen) in the bulk or at the surface, and act as donors. \cite{Panayotov2007,Peacock2003,Herklotz2011} Adsorption at the surfaces of the $\mathrm{TiO_2}$ pores may play a role as well,\cite{Kowalski2009} where atomic hydrogen tends to rest preferentially at the top of bridging oxygen atoms.\cite{Islam2011}. However, the diffusion of hydrogen atoms into the subsurface is significant,\cite{Aschauer2012}, and hydrogen diffuses extremely well into the bulk along the c-axis in rutile.\cite{Bates1979} The time constant of the approximately exponential response of $I(t)$ is $2\,\mathrm{min}$ for $c(\mathrm{H_2})=15\,\mathrm{ppm}$ and decreases rapidly as $c(\mathrm{H_2})$ is increased. The sensing is mostly reversible, except a slowly increasing current background, which we attribute to incomplete hydrogen removal.\\
In Figs. \ref{TiOx_Fig2}(a) and (b), the response of $I(t)$ to $\mathrm{H_2}$ pulses of 3 min. duration is shown for $V=10\,\mathrm{V}$ in the interval where electroforming increases the conductance. In response to the first hydrogen pulse, offered at $t = 1\,\mathrm{min.}$, $I(t)$ first \emph{decreases} and then begins to increase. The decrease, which is absent for voltages below the electroforming threshold, can be larger than one order of magnitude. For $c(\mathrm{H_2})< 100\,\mathrm{ppm}$, it dominates the overall behavior and the conventional sensing response is not observed. Only for $c(\mathrm{H_2})\geq 100\,\mathrm{ppm}$, a current peak starts to develop shortly after the decrease. The negative response gets faster as $c(\mathrm{H_2})$ is increased, while its amplitude shows no clear correlation with $c(\mathrm{H_2})$ and fluctuates among $I(t)$ traces under nominally identical conditions (not shown). Afterwards, the sample responds to hydrogen in the usual way \cite{Varghese2003,Paulose2006,Cerchez2013} in this time interval. Furthermore, for $c(\mathrm{H_2})\leq 30\,\mathrm{ppm}$, hydrogen sensing and electroforming appear approximately as a superposition, while for larger hydrogen volume fractions, no further increase of $I(t)$ due to electroforming can be observed after the first two hydrogen pulses. This suggests that exposure to hydrogen accelerates the electroforming.\\
We proceed by studying the response of $I(t)$ to $\mathrm{H_2}$ after $I_{max}$ has been reached, i.e. when phase 1 has disappeared. Again, an initial decrease of $I(t)$ in response to the first $\mathrm{H_2}$ pulse is observed, which can be very strong \cite{Supplement} and vanishes rapidly under subsequent $\mathrm{H_2}$ pulses. In addition to the increase of $I(t)$ in response to the $\mathrm{H_2}$ pulse, a simultaneous process is active that decreases $I(t)$ continuously in the presence of hydrogen.\cite{Supplement} This manifests itself in plateaus of $I(t)$ in between the $\mathrm{H_2}$ pulses that form steps towards a smaller current as time proceeds. While the positive response of $I(t)$ is essentially reversible, its accelerated decrease is irreversible  but can be reset by our initialization procedure.\\
This interdependence of hydrogen sensing and electroforming is interpreted in terms of an \emph{ad hoc} - extension of the single filament model.\cite{Strukov2008} In each phase j, the conductivity $\sigma_j(t)$ is assumed to be the sum of an oxygen vacancy - generated component $\sigma_j^{o}$ and a phase-independent component $\sigma_A(t)= \sigma_A(c(\mathrm{H_2}))\times[1-e^{-t/\tau_A}]$ due to hydrogen doping (mechanism A), see the insets in Fig. \ref{TiOx_Fig3}(a,d). The doping therefore tends to homogenize $\varepsilon (x)$ along the filament, leading to an increase of $\varepsilon_2$, the electric field in phase 2. Since the velocity $v_b (t) $ of $x_b (t)$ is determined by $\varepsilon_2$, hydrogen doping accelerates $x_b(t)$. Before $I_{max}$ is reached, the hydrogen doping accelerates the expansion of phase 2 and with it the increase of $I(t)$. After $I_{max}$ has been reached, the doping accelerates the diminution of phase 2, which gets replaced by the less conductive phase 3. This process is readily incorporated in the single filament model \cite{Strukov2008,Supplement}. In the simulations, a filament of $5\,\mathrm{\mu m}$ length and an oxygen vacancy drift mobility of $\mu(V_O)=10^{-15}\,\mathrm{m^2/(Vs)}$ are used. In Fig. \ref{TiOx_Fig3}(a), the result of such a calculation is shown for a $\mathrm{H_2}$-induced exponential increase of $\sigma_1 (t)$ and $\sigma_2 (t)$, with $\tau_A$ set to $200\,\mathrm{s}$. Both $I(t)$ and $v_b(t)$ increase during the hydrogen pulse, and $I(t)$ shows good qualitative agreement with the measured phenomenology. Likewise, $I(t)$ can be modeled after $I_{max}$ has been reached, see Fig. \ref{TiOx_Fig3}(b). Here, after a $\mathrm{H_2}$ pulse has finished, $I(t)$ has dropped below the value it would have reached without hydrogen exposure, and this difference decreases as phase 3 spreads out and $x_b$ approaches the Ti substrate.\\
In order to integrate the additional, initial decrease of $I(t)$ in response to $\mathrm{H_2}$ exposure, we postulate a co-existing mechanism by which atomic hydrogen decreases the conductivity (mechanism B). A decrease of $\sigma_1$ and $\sigma_3$, i.e., of phases composed of weakly reduced $\mathrm{TiO_2}$, in response $\mathrm{H_2}$ exposure has not been observed up to now and is not to be expected, considering that interstitial hydrogen acts as donor.\cite{Bjorheim2010,Panayotov2007,Peacock2003,Herklotz2011} We therefore assign mechanism B to the strongly reduced phase 2. A  microstructural explanation of mechanism B could be related to the work of Filippone et al. \cite{Filippone2009}, who suggested that a $\mathrm{H}$ atom  may form a complex with an oxygen vacancy, with the effect of localizing one of its two delocalized electrons. The effect would be a decrease of the electron density in the conduction band. Furthermore, since the binding energy of such a hydrogen-oxygen vacancy complex has been calculated to be $0.49\,\mathrm{eV}$ larger than that one of the oxygen vacancy, \cite{Filippone2009} this configuration is expected to be stable, in agreement with the observed irreversibility.
Figs. \ref{TiOx_Fig3}(c) and (d) show that the effects of mechanism B on $I(t)$ can be reproduced as well within the single filament model. Here, is is assumed that due to mechanism B, only $\sigma_2^{o}$ develops a dependence on $c(\mathrm{H_2})$ and experiences a nonrecurring exponential decrease according to $\sigma_2^{o}(t)= \sigma_{2,B}^{o}(c(\mathrm{H_2}))+[\sigma_2^{o}-\sigma_{2,B}^{o}]e^{-t/\tau_B}$ with time constant  $\tau_B=10\,\mathrm{s}$, while $\sigma_1$ and $\sigma_3$ remain unaffected. Different hydrogen concentrations are modeled by different final conductivities for all phases.\cite{Strungaru2015}

As hydrogen is offered, an initial decrease of $I$ is obtained which correlates with an acceleration of $x_b$ due to the additional increase of $\varepsilon_2$. A subsequent increase of $I$ shows that mechanism A starts to dominate again. Here, $v_b$ is larger than before the hydrogen pulse because of the increase of $\varepsilon_2$. After hydrogen is turned off before $I(t)$ has reached $I_{max}$, the current increases more slowly due to the growth of phase 2 or, depending on the hydrogen volume fraction, decreases, in qualitative agreement with the experimental studies, see Fig. \ref{TiOx_Fig2}(c). After $I(t)$ has passed $I_{max}$, the accelerated expansion of phase 3 during $\mathrm{H_2}$ exposure causes $I(t)$ to decrease below the value it would have reached if hydrogen were absent, see Fig. \ref{TiOx_Fig2}(d).\\
It has been recently established that electroforming leads to a transformation of $\mathrm{TiO_2}$ into the $\mathrm{Ti_4O_7}$ Magn\'{e}li phase inside filaments, \cite{Kwon2010} which can be thought of as a local recrystallization of reduced $\mathrm{TiO_{2-\delta}}$. We therefore conclude by stressing that the observed behavior is also consistent with the formation of  Magn\'{e}li phases $\mathrm{Ti_nO_{2n-1}}$ with integer n. Let us assume that phase 2 is composed of $\mathrm{Ti_4O_7}$. Since it is well established that $\sigma(\mathrm{Ti_4O_7})\approx 10^3\sigma(\mathrm{TiO_2})$, \cite{Bartholomew1969,Szot2011} it will have a larger conductivity than phase 1. Offering hydrogen may reduce $\mathrm{Ti_4O_7}$ further, i.e., to n=2 or 3, with smaller conductivities than $\mathrm{Ti_4O_7}$, \cite{Bartholomew1969,Szot2011} which would be an alternative explanation for mechanism B.

To summarize, the coexistence and interdependence of electroforming and hydrogen incorporation in titanium dioxide films has been demonstrated. Atmospheric hydrogen accelerates the electroforming via doping and a corresponding redistribution of the electric fields among the phases characterized by different oxygen deficiencies. In addition, a rapidly decreasing, nonrecurring and irreversible current component in response to hydrogen is observed. It is tentatively attributed to hydrogen - oxygen vacancy complexes, but can also be understood within a picture based on Magn\'{e}li phases. The measurements can be reproduced qualitatively by simulations based on the filament model. It is a task for the future to study the microscopic origin of these effects, their interplay with other atmospheric gases like, e.g., oxygen, which is known to be nontrivial even below the electroforming threshold,\cite{Cakabay2015} as well as the role of the pore surfaces vs. that one of the bulk material.

We thank U. Zimmermann for assistance during the measurements.

%

\end{document}